\documentclass[12pt, draftcls, onecolumn]{IEEEtran}
\usepackage{graphics, graphicx}
\usepackage{amssymb,amsmath}
\usepackage{amsmath}
\usepackage[rm]{subfigure}    
\usepackage{threeparttable}   
\usepackage[dvips]{color}
\usepackage{ams fonts}   
\usepackage{url}   
\usepackage[latin1]{inputenc} 
\usepackage{pifont}
\usepackage{algorithm}
\usepackage{cite}
\usepackage[dvips]{color}
\usepackage{lineno}
\usepackage{pstricks}
\usepackage{tikz}
\definecolor{gold}{rgb}{0.85,.66,0}

\newcommand{\GF}{\rm GF}
\colorlet{examplefill}{yellow!80!black}
\tikzstyle{every plot}=[prefix=plots/pgf-]
\tikzstyle{shape example}=[color=black!30,draw,fill=yellow!30,line width=.5cm,inner xsep=2.5cm,inner ysep=0.5cm]

\begin{document}
\title{Error-Correcting Codes for Reliable Communications in Microgravity Platforms}
\author{\IEEEauthorblockN{D\'ecio L. Gazzoni Filho, Taufik Abr\~ao, Marcelo C. Tosin, Francisco Granziera Jr}
\thanks{Electrical Engineering Department, State University of Londrina, P.O. Box 6001, Zip Code: 86051-990, Londrina, PR, Brazil, Email: dgazzoni@uel.br, taufik@uel.br, mctosin@uel.br, granziera@uel.br}}

\maketitle



\begin{abstract}
Rocket-launched microgravity platforms are employed to subject scientific experiments to a near-zero gravity environment. Experiments carried aboard these platforms may transmit telemetry data to ground stations as a hedge against vehicle recovery failure, a common occurrence in Brazilian missions. If the vehicle is lost, telemetry data is the only means of recovering scientific data collected in-flight; hence corruption of this data could jeopardize the experiment.

The authors were responsible for an experiment, the Platform for Acquisition of Acceleration Data (PAANDA), launched aboard the Brazilian Cum\~a II mission from the Alc\^antara Launching Center (CLA) in 2007. This instrument was conceived to assess the residual acceleration environment during the microgravity period of a microgravity platform. Theoretically it is capable of measuring accelerations with magnitudes of 1 $\mu$g. The instrument can also measure and store acceleration data of all flight phases of the vehicle, from launch through recovery.

Telemetry data is received at the CLA and Barreira do Inferno Launching Center (CLBI) ground stations. Our experiment experienced some data corruption, leading to loss of all telemetry information sent during the reentry period, as well as some from the microgravity period. For the second version of PAANDA, we sought to prevent such data corruption from happening again.

Traditionally, an error-correcting code for this channel would consist of a block code with very large block size to protect against long periods of data loss. Instead, we propose the use of digital fountain codes along with conventional Reed-Solomon block codes to protect against long and short burst error periods, respectively. We propose a model for the communication channel based on information extracted from Cum\~a II's telemetry data, and simulate the performance of our proposed error-correcting code under this channel model. Simulation results show that nearly all telemetry data can be recovered, including data from the reentry period, using coding rates well within the capacity of the vehicle's telemetry equipment.
\end{abstract}

\begin{keywords}
Microgravity Platform, PAANDA, Error-Correcting, Digital Fountain codes
\end{keywords}

\section{Introduction}
Microgravity platforms are used to create environments that are nearly acceleration-free. Usually, these platforms are launched by a sounding rocket which follows a sub-orbital parabolic trajectory. During its flight, the platform enters a free-fall state and cancels its own rotations, creating the microgravity environment. This microgravity state lasts for a few minutes before the platform reaches the upper atmosphere, where aerodynamic drag starts to act on the platform. Experiments carried aboard the platform may transmit telemetry data back to ground stations, to protect against data loss in the event that the vehicle is unable to be recovered.

The Platform for Acquisition of Acceleration Data (PAANDA) instrument was conceived to assess the residual acceleration environment during the microgravity period of a microgravity platform \cite{PAANDA_RELAT}. Theoretically it is capable of measuring accelerations with magnitudes of 1 $\mu$g. The instrument can also measure and store acceleration data throughout all phases of the flight, including the vehicle's propelled phase.

The first PAANDA prototype has flown in the Cum\~a II mission. It was launched from the Alc\^antara Launching Center (CLA) in July 19, 2007, and remained in a microgravity state for $122$ s. PAANDA's telemetry data was received (with errors) by the CLA ground station, as well as the Barreira do Inferno Launching Center (CLBI) ground station. Unfortunately, vehicle recovery is a risky operation which was not carried out successfully for this mission, as well as the previous mission (Cum\~a I). Hence, a significant portion of measurements were lost, including all of the data during the atmosphere re-entry phase, when a complete communications blackout occurred. For the benefit of the scientific experiment being carried out, we decided to design and implement an error correcting code which might be able to recover almost all data, even in the presence of long-lasting blackouts. The resulting error-correcting code design is chronicled in this paper.

In order to validate our design, we present a channel model obtained from the scarce data that is available from the Cum\~a II mission, and perform simulations based on this model to assess the performance of the proposed error-correcting code, which is based on the concatenation of conventional Reed-Solomon block codes with the more recent class of Digital Fountain codes. Due to these promising simulation results, this code will be implemented in the next version of the instrument, called PAANDA II, which is currently under development and whose launch is yet to be scheduled.

\subsection{A brief description of PAANDA}

The Platform for Acquisition of Acceleration Data (PAANDA) was designed to be one of the experiments carried aboard a rocket launched microgravity platform. This instrument mainly consists of three high sensitivity pendular accelerometers mounted on an orthogonal frame, data conversion electronics and an embedded computer for system management. Mission requirements dictate that the system must measure a wide range of accelerations; PAANDA has two operating scales to ensure measurements are performed at the appropriate resolution in spite of this wide dynamic range. The high magnitude scale operates from an absolute acceleration of 18 g down to 1.05 g, when it automatically switches to the high resolution scale, which can measure accelerations as low as 1 $\mu$g.

Three identical accelerometers (Honeywell QA2000-010) are employed. This model is a pendular force-balance accelerometer with closed-loop control \cite{qa2000}, which has been successfully used in space missions before \cite{hassi}. The instrument generates acceleration data at a rate of 10 samples per second. The system also measures temperature data from various important components of its data acquisition circuits, including the sensors. The instrument's central embedded computer encodes and packs this data using a basic ASCII code strategy. The generated packed data includes a checksum and synchronization information. The data packets are transferred through a serial RS-422 link to the vehicle's telemetry equipment, which transmits it to the ground stations.

The retrieved system information, such the accelerometer operating scale and temperatures, are latter used to correct the acceleration data by placing it into polynomial models of the accelerometers and other sensitive components of the data acquisition subsystem. The accelerometer offsets are also corrected using calibration data previously obtained from experimental procedures prior to flight.

\section{The Problem and Proposed Solution Approach}
\subsection{Typical Launching Scenario and Problem Description}

The communications system considered in this paper is composed of a transmitter aboard the microgravity platform, and two receivers at the ground base stations of Natal and Alc\^antara. Users of the microgravity platform supply data to be transmitted through a serial RS-422 interface with a maximum bit rate of 76.8 kbps. This interface uses one start bit and one stop bit for each transmitted byte, so that the effective data rate is 7680 bytes/s. Other than that, the only known information about the system is that an error-correcting code is inserted by the telemetry system that corrects a single bit error per byte. We hypothesize this is a (15,11) Hamming code shortened to (12,8), resulting in a raw bit rate of 115.2 kbps, which is a standard rate for a serial link.

\begin{figure}[!htbp]
\centering
\includegraphics[width=0.45\textwidth]{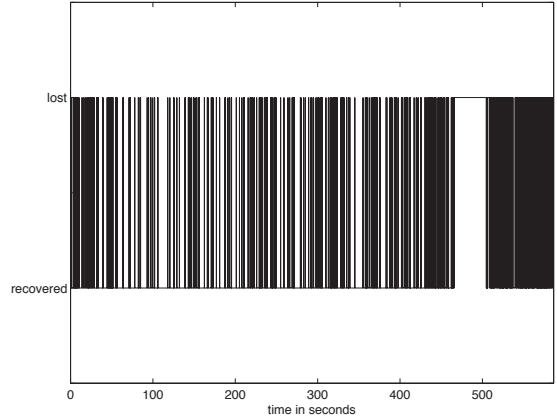}
\vspace{-3mm}
\caption{Lost and received packets over time. Reentry period is around 465 to 504 sec.}
\label{fig:lostpackets}
\end{figure}

The rocket is launched from Alc\^antara, which initially is the only station to receive data from the platform (due to the Earth's curvature, at launch the Natal station has no line of sight to the transmitter). After reaching a certain altitude, the Natal station comes online and remains so until the rocket re-enters the atmosphere and drops below the altitude necessary to maintain line of sight. Based on telemetry data from the Cum\~a II mission on which PAANDA has flown, we observe that the telemetry system's BER is acceptable during most of the flight, and we conjecture (with support from the results of Section \ref{sec:chan_model} and \ref{sec:results}) that the use of a standard, low-rate block code such as (255,223) Reed-Solomon could achieve arbitrarily low BERs. However, during the reentry period, BER grows considerably and we observe a complete loss of data transmitted during this period. Figure \ref{fig:lostpackets} illustrates packet loss over the time of the flight.

\subsection{Design approaches}
Due to the scientific value of experiment data gathered during the reentry phase, we investigate error correction code design methodologies that could salvage the data lost in this period. In Section \ref{sec:conv_rx_topology} we investigate a conventional receiver topology based on block error-correcting codes with very large block sizes and find it lacking, leading us to propose in Section \ref{sec:proposed_topology} an innovative topology based on a recent class of error-correcting codes called digital fountain codes.

\subsubsection{Conventional receiver topology} \label{sec:conv_rx_topology}
The classical approach to designing an error correcting code for the described channel would call for a block code with good burst error correcting capability, such as Reed-Solomon (RS) codes \cite{ReedSolomon}. We briefly review these codes here.

Let $m$ be the number of bits in a symbol; most RS codes typically take $m = 8$. We work in the finite field $\GF(2^m)$. Let $n = 2^m - 1$ be the order of the multiplicative group of $\GF(2^m)$; this is the maximum block size of the code\footnote{Shorter blocks can be constructed using the technique of \emph{shortening}, but we will not consider shortened codes in this explanatory section.}. An $(n,n-k)$ RS code is given $n-k$ symbols as input, and output $k$ parity symbols by oversampling a polynomial related to the input symbols; the codeword is composed of the original $n-k$ symbols plus the $k$ parity symbols. This code can correct up to $\lfloor k/2 \rfloor$ errors in unknown positions of the code words, or $k$ errors in known positions (termed \emph{erasures}). A typical code is the RS $(255,223)$ code, which can correct up to 16 errors and 32 erasures.

Although larger block sizes can be used (by taking e.g. $m = 16$, so that codes of size up to 65535 symbols are possible), the computational complexity, both of finite field arithmetic and the code itself, increases rapidly. A more efficient solution, which achieves similar performance in certain situations, is to retain a small block size, while \emph{interleaving} the contents of consecutive blocks. For instance, an interleaved $(2550,2230)$ RS code can be constructed by interleaving 10 blocks of a $(255,223)$ ordinary RS code. For this interleaved code, not every error pattern of 160 errors can be corrected, but in the particular case of a single, long burst error, the interleaved code performance is identical to that of a larger RS code with no interleaving. There are, however, some drawbacks to the use of long block size codes: first, all input data must be known before the codewords are calculated, incurring an initial transmission latency which increases linearly with the block size; second, losing even a single block implies in the loss of a large amount of data. In a scenario with wide noise variance, a short block size code could recover some data during periods of less noise, whereas a long block size code would average the noise over the length of the block and lose all of its data.

\subsection{Channel model} \label{sec:chan_model}

In order to validate the proposed error-correcting code through simulation, we attempted to model the communication channel over which telemetry data is transmitted. Since very little is known about the topology of the system, we rely on empirical results gathered from the telemetry data of the PAANDA experiment, launched in a previous mission (Cum\~a II). PAANDA employed a simple format for data encoding, with a start symbol, the data itself, a CRC-16 checksum of the data and a stop symbol. Some redundancy was added to the data by representing 8-bit symbols as two hexadecimal digits in ASCII. Each packet has a fixed size of 44 bytes.

Unfortunately, in this mission the vehicle wasn't recovered, making it impossible to reconstruct the transmitted packets in order to compare to the received packets at the symbol level. As such, the only remaining choice is to assess the correct reception of complete packets, according to whether the packet checksum matched or not. We plot the state of received packets over time in Figure \ref{fig:lostpackets}.

Four distinct regions are highlighted in this plot. In the first region (the first $\approx 63$ secs), the vehicle is in full acceleration for the first $\approx 40$ secs (complicating the task of antenna tracking) and only the Alc\^antara ground station is receiving data, so the packet error rate (PER) is fairly high. In the second region (from $\approx 67$ to $\approx 465$ secs in Fig. \ref{fig:lostpackets}), PER decreases considerably due to the Natal station coming online, in addition to a more predictable vehicle trajectory. This region includes the microgravity period of the flight. The third region is characterized by a complete communications blackout (the next $\approx 39$ secs) due to the vehicle re-entering the atmosphere. The fourth region encompasses the end of the flight (the last $\approx 82$ secs), while the vehicle falls down back to the earth until communications synchronism is lost. It has similar characteristics to the first region, as far as data transmission is concerned, although the PER is somewhat higher.

Our channel model considers these four regions separately. In light of the fact that the only available information is from a single flight, not much can be inferred about the variance of transition points between each region of the flight. As such we will model these transitions as purely deterministic and occurring at exactly the same time as they were recorded during the Cum\~a II flight.

Given the line of sight (LOS) characteristics of the simplex telemetry communication over a (essentially) free-space non-guided propagation channel, the AWGN channel model assumption is suitable for the first, second and fourth regions of the vehicle's flight. A simple non-coherent post-detection antenna selection diversity (from the Natal and Alc\^antara ground stations) is employed to select data detected from the two non-coordinated receivers over the complete flight. The decision rule used to obtain the packets illustrated in Fig. \ref{fig:lostpackets} was simply based on the checksum values for each received packet at the Natal and Alc\^antara ground stations.

Modeling the re-entry (third) region is straightforward, as it consists of a complete communications blackout. Our model consists of dropping all packets during this period. For the remaining regions, we used an AWGN channel model with different SNR values for each region. We simulated the setup found in the original PAANDA experiment, with 44-byte packets which are considered lost in case of a CRC-16 mismatch, meaning a single bit error leads to a lost packet. A (12,8) Hamming code was added before transmission, to provide the claimed 1 bit per byte error correcting capability of the telemetry system. The simulation was performed with varying levels of SNR, seeking a value that produces an average packet loss which coincides with the observed packet loss in the Cum\~a II mission\footnote{Given that a single data point was available, the maximum likelihood estimate for the average packet loss is given by this data point.}. Using this methodology, we estimate the SNR at the first, second and fourth flight regions (Table \ref{tab:SNR_regions}). For the blackout region, we fix the SNR as low as possible, i.e., $\approx -1.6dB$, since the entire data in this period was lost.

\begin{table}[htbp!]
  \centering
  \caption{Estimated SNR for the four rocket flight regions of the Cum\~a II mission}\label{tab:SNR_regions}
  \begin{tabular}{c|cccc}
    \hline
    &\multicolumn{4}{c}{\textbf{Flight Regions}}\\
      \cline{2-5}
   Estimated & 1 & 2 & 3 & 4 \\
    \cline{2-5}
    SNR [dB] & $4.75$ & $5.10$ & $-1.60$ & $3.65$ \\
    \hline
  \end{tabular}
\end{table}

\section{Proposed topology using digital fountain codes} \label{sec:proposed_topology}

We propose the use of rateless erasure codes, also known in the literature as \emph{digital fountain} codes. Some realizations of this concept are Tornado codes \cite{Tornado}, Online codes \cite{Online}, and LT codes \cite{LT}.

Erasure codes are designed to work in channels where data is either correctly received or lost; it is assumed that there are no unflagged errors. The data is laid out as a series of packets. Most of these codes rely on the same basic idea: construct codewords from random linear combinations of different data packets (for binary data, this is achieved by the use of the XOR operation), and after the reception of enough codewords, use linear algebra to extract the original data packets from the codewords. Generally, each type of rateless erasure code has an associated statistical distribution which is sampled to construct the codewords. Typically, statistical distributions are chosen so as to produce sparse linear combinations (i.e. few data packets are included in each codeword), while still retaining enough information for decoding. Obviously, sparseness leads to a decreased computational complexity in the encoder, but more interestingly, decoding (which must perform a task functionally similar to Gaussian elimination) can also be performed efficiently. To illustrate, in a data set composed of $n$ packets, efficient decoding algorithms generally have an asymptotic complexity of $O(n)$ or $O(n \log n)$ versus $O(n^3)$ for Gaussian elimination.

However, such sparsity comes at a cost: most of these codes require receiving approximately 105 to 110\% as many codewords as there are data packets before a full-rank matrix can be constructed. On the other hand, a large dense random matrix requires only a constant number of codewords beyond the minimum required for decoding (on average, less than two \cite{BlakeStudholme_06, BlakeStudholme_10}), increasing the chance of correct decoding given the same amount of correctly received codewords. Moreover, most of these codes assume that the complete data set is available at the start of the encoding process, which is not the case for a telemetry application such as the one described in this paper. It follows that many of the assumptions made in deriving the statistical distributions that are sampled to construct codewords are invalid. If increased encoding and decoding complexity is acceptable (which is the case in our application), then a natural choice is the use of dense random matrices for encoding \cite{Tirronen_09, Bogino_07}, and Gaussian elimination for decoding \cite{BlakeStudholme_06}. This is the choice of code employed in our design.

\section{Numerical results} \label{sec:results}

In this section, we use the channel model from Section \ref{sec:chan_model} to simulate possible error-correcting codes to be implemented in the PAANDA II instrument.

The packet format for PAANDA II was overhauled, due to additional data being monitored by PAANDA II compared to PAANDA, in addition to the desire for a larger block size in order to increase the effectiveness of block error-correcting codes. The new packet format has a length of $\approx 3300$ bytes and carries 1 second of instrumentation data, compared to 44 bytes for 100 ms of instrumentation data for PAANDA. This corresponds to a rate of 33 kbps over the RS-422 telemetry link, out of a maximum rate of 76.8 kbps. Error-correcting codes were designed to make use of the remaining 43.8 kbps.

We simulated four possible encoding schemes, covering a good range of tradeoffs for code design criteria:
\begin{enumerate}
\item uncoded (for comparison purposes only);
\item (255,111) Reed-Solomon (RS) code;
\item (255,223) Reed-Solomon code concatenated with a rate-1/2 Digital Fountain (DF) code;
\item (255,191) Reed-Solomon code concatenated with a rate-4/7 Digital Fountain code.
\end{enumerate}

We note that, for the uncoded case, large block sizes are not required, and indeed only serve the purpose of increasing error rates. As such, the original 3300-byte packet were split in a hundred 33-byte packets and transmitted independently from each other, so as to ``level the playing field'' somewhat. However, for such a small packet size, packet overhead is non-negligible, so we added a synchronization header (3 bytes), timestamp (2 bytes) and CRC-32 checksum (4 bytes) for a total of 42 bytes per packet.

In addition to SNRs for the channel model as estimated in Table \ref{tab:SNR_regions}, we performed simulations with reduced SNRs (by 0.5 dB, 1.0 dB , 1.5 dB and 2.0 dB) for the first, second and fourth flight period, while still assuming a complete blackout during the reentry period. This is in order to test the resiliency of the proposed error-correcting codes to deteriorated channel conditions.

\begin{figure}[!htbp]
\centering
\includegraphics[width=0.6\textwidth]{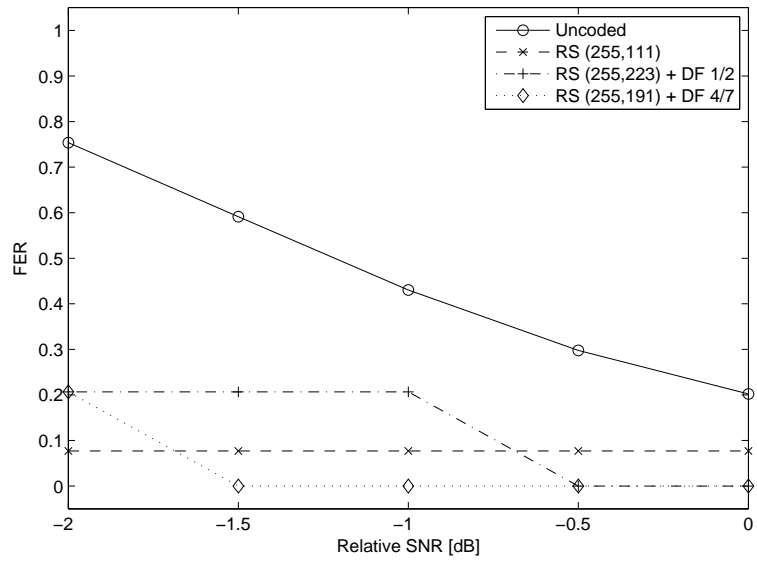}
\vspace{-3mm}
\caption{FER for the complete flight, considering SNR reduction regarding the reference SNR from Table \ref{tab:SNR_regions}.}
\label{fig:Overall_FER}
\end{figure}

\begin{figure}[!htbp]
\centering
\includegraphics[width=0.65\textwidth]{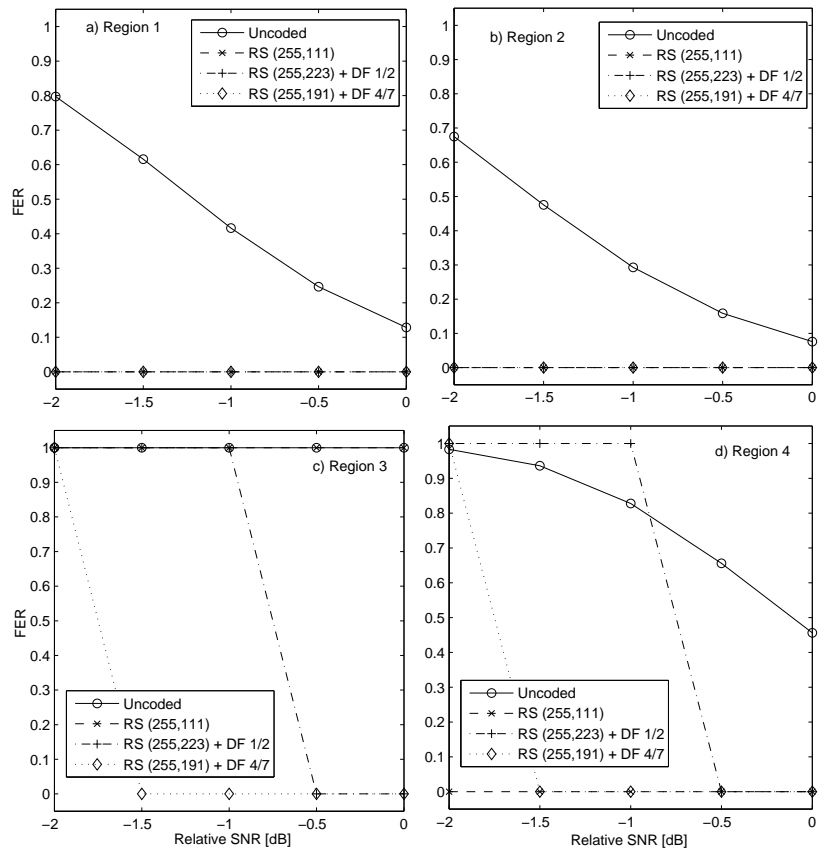}
\vspace{-3mm}
\caption{FER for each region of the flight, considering SNR reduction regarding the reference SNR from Table \ref{tab:SNR_regions}.}
\label{fig:FER_regions}
\end{figure}

We performed Monte-Carlo simulations of all four encoding schemes, each of them under the five described SNR scenarios, for a total of 20 combinations. Each combination was simulated with 250 trials and results were averaged. Simulation results are shown in Figures \ref{fig:Overall_FER} and \ref{fig:FER_regions}. In Figure \ref{fig:Overall_FER} we plot frame error rates (FER) for the complete flight, while in Figure \ref{fig:FER_regions} we break down the contributions of each flight region.

Results confirm that lack of coding leads to a significant frame error rate, as high as 75\% in one of the simulated scenarios. Reed-Solomon coding alone is capable of correcting all errors for all chosen values of SNR, except for the blackout region (Fig. \ref{fig:FER_regions}.c), as expected. The proposed concatenation of a Reed-Solomon (255,223) code with a rate-1/2 digital fountain code allows complete data recovery (0\% FER) for the estimated SNR, and even tolerates an SNR reduction of 0.5 dB. However, further reductions in SNR lead to complete data loss (100\% FER) during the blackout region, even extending into the fourth region, an unfortunate side effect of unsuccessful decoding of the digital fountain code. Data recovery during the first and second flight regions is not affected, though.

Successful decoding of the digital fountain code requires enough recovered packets to build a full-rank decoding matrix. This can be achieved either by transmitting more packets (through increasing the digital fountain code rate), or reducing the amount of packets lost to unrecoverable errors (through increasing the Reed-Solomon code rate). Hence transmission bandwidth must be split between the Reed-Solomon and digital fountain codes, giving rise to a spectrum of performance tradeoffs. Inspection of intermediate simulation results revealed that the Reed-Solomon code rate was inadequate given the amount of errors it is expected to correct under lower SNR scenarios. On the other hand, the digital fountain code could tolerate a minor reduction in rate. As such, we simulated a different point in the design space of these concatenated codes; namely, a (255,191) RS code concatenated with a rate-4/7 digital fountain code. Figures \ref{fig:Overall_FER} and \ref{fig:FER_regions} show that this different set of parameters can tolerate a further 1 dB reduction in SNR while maintaining complete data recovery, down to 1.5 dB below the estimated SNR of the Cum\~a II mission. Although it was unable to fully recover the data after a further 0.5 dB drop in SNR, we conjecture that there are yet another point in the design space which could achieve 0\% FER.

We stress, however, that the originally proposed parameters are more than adequate for the estimated SNR of this channel, recalling its ability to achieve 0\% FER for every one of 250 trials of the the carried out Monte-Carlo simulations.

\section{Conclusion}
The results of Section \ref{sec:results} indicate that the conventional approach to error-correcting code design is ineffective in the face of communication blackouts. The use of a newer class of codes, rateless erasure codes or digital fountain codes, was shown to be more promising in this application.

An area of interest for further research is the design of efficient digital fountain codes for streaming applications. The literature on digital fountain codes concentrates on the case where the complete data set is known \emph{a priori}, which was shown to be an invalid assumption in our application. This required resorting to a less efficient code with encoding complexity that grows quadratically with data set size, and decoding complexity that grows cubically. Seeing as log-linear and even linear complexities are achieved for the \emph{a priori} scenario, it is not a stretch to imagine that they could be achieved in a streaming application as well.


\section*{Acknowledgment}
This works was partially sponsored by the Brazilian Space Agency Microgravity Program, grant 012/2004 and in part by the National Council for Scientific and Technological Development (CNPq) of Brazil under Grant 303426/2009-8.



\end{document}